\documentclass[12pt]{iopart}
\usepackage{graphicx}
\usepackage[colorlinks,citecolor=blue,linkcolor=red]{hyperref}

\newcommand{\braket}[2]{\left<#1|#2\right>}
\newcommand{\ketbra}[2]{\vert #1 \rangle \langle #2 \vert}

\newcommand{\ket}[1]{| #1 \rangle}
\newcommand{\bra}[1]{\langle #1 |}

\newcommand{\blue}[1]{\textcolor{blue}{#1}}

\begin{document}

\title[Cluster state generation in one-dimensional Kitaev honeycomb model via shortcut to adiabaticity]{Cluster state generation in one-dimensional Kitaev honeycomb model via shortcut to adiabaticity}

\author{Thi Ha Kyaw$^1$}
\ead{thihakyaw@u.nus.edu}
\author{Leong-Chuan Kwek$^{1,2,3,4}$}
\ead{kwekleongchuan@nus.edu.sg}
\vspace{10pt}

\address{$^{1}$%
	Centre for Quantum Technologies, National University of Singapore, 3 Science Drive 2, Singapore 117543, Singapore%
}

\address{$^{2}$%
	MajuLab, CNRS-UNS-NUS-NTU International Joint Research Unit, UMI 3654, Singapore
}

\address{$^{3}$%
	Institute of Advanced Studies, Nanyang Technological University, 60 Nanyang View, Singapore 639673, Singapore%
}

\address{$^{4}$%
	National Institute of Education, Nanyang Technological University, 1 Nanyang Walk, Singapore 637616, Singapore%
}
\vspace{10pt}
\begin{indented}
\item[]Updated on \today
\end{indented}

\begin{abstract}
We propose a mean to obtain computationally useful resource states also known as cluster states, for measurement-based quantum computation, via transitionless quantum driving algorithm. The idea is to cool the system to its unique ground state and tune some control parameters to arrive at computationally useful resource state, which is in one of the degenerate ground states. Even though there is set of conserved quantities already present in the model Hamiltonian, which prevents the instantaneous state to go to any other eigenstate subspaces, one cannot quench the control parameters to get the desired state. In that case, the state will not evolve. With involvement of the shortcut Hamiltonian, we obtain cluster states in fast-forward manner. We elaborate our proposal in the one-dimensional Kitaev honeycomb model, and show that the auxillary Hamiltonian needed for the counterdiabatic driving is of $M$-body interaction. 
\end{abstract}

%
%
%
%
%

\section{Introduction}
A quantum computer promises efficient processing capability for certain computational problems in contrast to current classical computer \cite{papageorgiou2013,ronnow2014,heim2015}. In order to build and design efficient quantum circuitry that outperforms its classical counterpart, it is essential to exploit the unique quantum mechanical features that optimize and enhance computations. It has been well-accepted and demonstrated that quantum entanglement, one of the main pillars of quantum information processing, gives rise to an important resource for quantum speed-up \cite{shor1994,grover1997}. Typically, the input states to a typical quantum circuit are not entangled and of the form $\ket{0}\otimes\ket{0}\otimes\cdots\otimes\ket{0}$, where $\ket{0}$ represents the ground state of some two-level system, and the entanglement needed for the quantum computation must be generated within the circuit itself. There exists an alternative paradigm, where the desired quantum gate operations are obtained through single-particle projective measurements on some highly entangled resource states or cluster states \cite{van2006}. This is known as the measurement-based quantum computation (MBQC) \cite{raussendorf2001,raussendorf2009,kwek2012}. The caveat is that one needs to prepare highly entangled states before the MBQC algorithm begins. Moreover, the preparation of these resources should preferably not be done through two-qubit entangling gate operations on some available physical qubits since one can simply perform the standard quantum circuit algorithm with entangling gates. Consequently, a preferred way to prepare and obtain these resources is to consider physical systems whose ground states are precisely these entangled resources and obtaining the states through cooling the system to its ground state. One such resource for the MBQC is the cluster state \cite{briegel2001} which is the ground state of spin-1/2 particles with $k$-body interactions where $k\geq 3$ \cite{raussendorf2005,jennings2009}. Unfortunately, cluster states cannot be obtained as a unique ground state of any Hamiltonian with only two-body interactions \cite{nielsen2006}.

There are some proposals in the literature to skirt around this obstacle. One such proposal \cite{raussendorf2001,mandel2003controlled,albarran2017} involves the creation of cluster states with only nearest-neighbor Ising-type interactions through precise control over the time evolution.  In \cite{kyaw2014measurement}, the present authors have also proposed an adiabatic scheme in which  one could obtain cluster states without the need to cool a system down to a very low temperature. The essential ingredients of the proposal are as follow. First, the systems are cooled down to its unique ground state, which is not a cluster state, with a large energy gap.  Second, some system parameters are then tuned adiabatically to reach to desired cluster states. These new states could have a much smaller energy gap compared to the initial one present in the system. Thanks to the inherent stabilizers symmetry of the system, the desire ground state can be protected from noise or  fluctuations in the parameter space by the finite adiabatic switching rate \cite{kyaw2014measurement}. 

In this report, we present a mean to obtain cluster states through a technique called shortcut to adiabaticity or sometimes the transitionless quantum driving algorithm \cite{demirplak2003,demirplak2005,berry2009}. Its extensive applications range from frictionless dynamics in Bose-Einstein condensates \cite{del2011fast}, rapid displacement of ions in phase space \cite{an2016shortcuts}, fast holonomic quantum computing \cite{zhang2015fast}, shortcuts to non-Abelian braiding in topological quantum information \cite{karzig2015shortcuts}, to assisted tracking of many-body states \cite{del2012assisted,damski2014counterdiabatic}, etc. We refer to this review article \cite{torrontegui2013} and references therein to have an overview of the entire field. And, an interesting proof of adiabatic theorem for quantum spin systems \cite{bachmann2017adiabatic} is also recommended. Unlike the previous proposal \cite{del2012assisted}, we are mainly interested in the ground state adiabatic passage of a many-body system to get useful cluster states needed in MBQC. More importantly, the instantaneous time-evolving state always remains in the same eigen-subspace, and it does not need to cross over any quantum critical point.

The paper is organised as follows. Section \ref{general_setup} outlines the general theoretical overview of the proposal.  Section \ref{Kitaev_honeycomb} derives the auxillary Hamiltonian needed to achieve transitionless quantum driving algorithm. We also present numerical results and discuss their implications. Finally, we provide some comments on the scheme and discuss the feasibility of the proposal in section \ref{conclusion}.

\section{General setup}\label{general_setup}
\subsection{Adiabatic approach}\label{sec:adiabatic}
In our model, each logical qubit of the cluster state is composed of several spin-1/2 particles. We next consider a Hamiltonian in which the stabilizers of the required cluster state commutes with it. The initial state is the ground state of this system and it has a sufficiently large energy gap. We note that the initial state is not a resource for MBQC, since it lies outside the subspace where logical qubits are encoded. At the end of the adaibatic evolution, the final state is a cluster state of logical qubits that can then be converted into a cluster state of spin-1/2 particles via single-qubit measurements. 

We begin by encoding each $j$th logical qubit of the cluster state in $n$ spin-1/2 particles:
\begin{equation}
\ket{0}_j = \otimes _{m=1}^n \ket{\uparrow}_{j,m},
\hspace{0.8cm}
\ket{1}_j = \otimes _{m=1}^n \ket{\downarrow}_{j,m}.
\end{equation}
where $\ket{\uparrow}_{j,m}$ ($\ket{\downarrow}_{j,m}$) is the eigenstate of the Pauli operator $\sigma ^z_{j,m}$ with the eigenvalue $+1$ ($-1$).
These logical states are stabilized by operators $\{ \sigma ^z_{j,1}\sigma ^z_{j,m}\}$, i.e., the logical states are common eigenstates of the stabilizer operators with eigenvalue $+1$. Pauli $X$ and $Z$ operators of the $j$th logical qubit are
\begin{equation}
X_j = \prod _{m=1}^n \sigma ^x_{j,m}
\hspace{0.5cm} \textrm{and} \hspace{0.5cm}
Z_j = \sigma ^z_{j,1}.
\end{equation}

The cluster state is the common eigenstate with eigenvalue $+1$ of cluster-state stabilizers \cite{raussendorf2001,raussendorf2009,kwek2012}
$
S_j = X_j\prod _{i\in nb(j)}Z_i = \prod _{m=1}^n \sigma ^x_{j,m} \prod _{i\in nb(j)} \sigma ^z_{i,1},
$
where $nb(j)$ stands for the set of nearest neighboring logical qubits of the $j$th logical qubit.
At the physical qubit level, the cluster state is stabilized by $\{S_j\} \cup \{\sigma ^z_{j,1}\sigma ^z_{j,m}\}$. We note that a product of stabilizers is also a stabilizer. Hence, cluster-state stabilizers can then be recast as
$
S_j^{\{ m_{j,i} \}} = S_j\prod _{i\in nb(j)}\sigma ^z_{i,1}\sigma ^z_{i,m_{j,i}} 
= \prod _{m=1}^n \sigma ^x_{j,m} \prod _{i\in nb(j)} \sigma ^z_{i,m_{j,i}},
$
where $\{ m_{j,i} \}$ is a string of numbers satisfying $1\leq m_{j,i}\leq n$.
In summary, if a state is stabilized by $\{S_j^{\{ m_{j,i} \}}\} \cup \{\sigma ^z_{j,1}\sigma ^z_{j,m}\}$ for any choice of $\{ m_{j,i} \}$, the state is the cluster state.
This cluster state of logical qubits can be converted into a cluster state of physical qubits by measuring $\sigma^x$ of arbitrary $n-1$ physical qubits of each logical qubit.
Therefore, this cluster state of logical qubits is a universal resource for the MBQC.

The adiabatic cluster-state scheme \cite{kyaw2014measurement} is performed by considering a system of $N\times n$ spin-1/2 particles under the Hamiltonian
\begin{equation}
H_0=H_s+\lambda V, \label{HG}
\end{equation}
where $\displaystyle H_s = \sum _{j=1}^N \sum _{m=1}^n \left(-J \sigma ^z_{j,m} \sigma ^z_{j,m+1}\right),$ assuming the periodic boundary condition $\sigma ^z_{j,n+1} = \sigma ^z_{j,1}$, and $J$ is the nearest neighbour coupling constant of Ising types. Here, $V$ denotes some two-body interactions that satisfy the constraints below.
\begin{enumerate}
	\item $V$ commutes with a set of cluster-state stabilizers $\{S_j^{\{ m_{j,i} \}}\}$ corresponding to a choice of $\{ m_{j,i} \}$, \&
	\item Non-zero interaction strength $\lambda$ lifts the ground states degeneracy, resulting the system $H_0$ to a unique ground state with a finite energy gap above it.
\end{enumerate}

Indeed, whenever we find a physical system that is in the form of (\ref{HG}) and satisfies the two constraints above, we are able to get around \cite{kyaw2014measurement} the no-go theorem \cite{nielsen2006} and get cluster states with just two steps. First, we cool the system with a nonzero $\lambda$ to its ground state. Second, we adiabatically switch off $\lambda$. In the adiabatic limit, the final state is the cluster state of logical qubits up to some single-particle Pauli operations. This protocol relies on the set of cluster-state stabilizers $\{S_j^{\{ m_{j,i} \}}\}$ that are conserved quantities for any value of $\lambda (t)$, i.e., $[ H_0, S_j^{\{ m_{j,i} \}} ]=0$, $\forall \lambda, t$. We also remark that $H_s$ commutes with $S_j^{\{ m_{j,i} \}}$. Hence, the unique ground state of $H_0$ for any nonzero $\lambda$ is the common eigenstate of cluster-state stabilizers, with corresponding eigenvalues $\{s_j^{\{ m_{j,i} \}}\}$, where $s_j^{\{ m_{j,i} \}} = +1$ or $-1$. Therefore, if the initial state is the ground state with a nonzero $\lambda$, the adiabatic theorem ensures the final state is still a common eigenstate of cluster-state stabilizers with the same eigenvalues.

When $\lambda$ adiabatically approaches to zero, the energy gap between the ground and first-excited states vanishes, which usually implies one has to slow down the rate of change of $\lambda$ to avoid any inadvertent excitation. Fortunately, in the degenerate subspace, i.e., the logical subspace, the cluster state is the only state with eigenvalues $\{s_j^{\{ m_{j,i} \}}\}$. Similarly, the ground state at $\lambda\neq 0$ is the only state with eigenvalues $\{s_j^{\{ m_{j,i} \}}\}$. Therefore, the transitions between the ground state and other states lifted from the degenerate subspace are forbidden; i.e., one does not have to slow down the rate of change of $\lambda$, even though there exists a vanishing energy gap, when $\lambda \rightarrow 0$ \cite{kyaw2014measurement}. In the following, we will focus on how to speed up the adiabatic process above by deploying the transitionless quantum driving algorithm.

\subsection{Shortcut to adiabaticity approach}\label{sec:shortcut}
Under the adiabatic evolution with a time-dependent Hamiltonian $H_0 (t)$ (\ref{HG}), a quantum system in its $n$-th eigenstate would remain in the same eigenstate as it undergoes time evolution. And, the instantaneous state is given by 
\begin{equation}
	\ket{\psi_n (t)}= \exp \left\lbrace -i\int_0 ^t dt' E_n (t') -\int_0 ^t dt' \braket{n(t')}{\partial_{t'}n(t')} \right\rbrace \ket{n(t)},\label{adiabatic_approx}
\end{equation}
where $\ket{n}(t)$ satisfies the Schr\"odinger equation $H_0 (t)\ket{n(t)}=E_n (t)\ket{n(t)}$. The first component in the exponent comes from the dynamical contribution and the second one is from the geometric contribution or the Berry phase \cite{berry1984}. By following the argument of Berry in \cite{berry2009}, we would like to find a new Hamiltonian $H(t)$ satisfying the Schr\"odinger equation ($\hbar=1$) $i\partial_t \ket{\psi_n (t)}=H(t) \ket{\psi_n (t)},$ which can be rewritten as 
$
	i\partial_t U(t)=H(t) U(t).
$
It follows that 
\begin{equation}
	H(t)=(i\partial_t U(t))U^\dagger (t),\label{H}
\end{equation}
where $U(t)$ follows the same adiabatic trajectory generated by the original Hamiltonian $H_0 (t)$. We know
\begin{eqnarray}
	U(t) &=& \sum_n \ketbra{\psi_n (t)}{n(0)}\nonumber\\
	&=& \sum_n \exp \left\lbrace -i\int_0 ^t dt' E_n (t') -\int_0 ^t dt' \braket{n(t')}{\partial_{t'}n(t')} \right\rbrace \ketbra{n(t)}{n(0)},
\end{eqnarray}
and similarly for $U^\dagger (t)$. After a few steps of algebra from (\ref{H}), we arrive at
\begin{eqnarray}\label{full_hamiltonian}
	H(t) &=& H_0 (t) + H_1(t) \\
	&=& \sum_n \ket{n(t)}E_n (t) \bra{n(t)}+i\sum_n \left(\ketbra{\partial_t n(t)}{n(t)}-\braket{n(t)}{\partial_t n(t)}\ketbra{n(t)}{n(t)} \right).\nonumber
\end{eqnarray}

The simplest example one can apply to this algorithm (\ref{full_hamiltonian}) is a two-level system in a magnetic field described by the Hamiltonian
\begin{equation}\label{two_level_hamiltonian}
	H_0 ^{\rm{2L}} (t) = \vec{h}(l(t)) \cdot \vec{\mathbf{\sigma}},
\end{equation}
where $\vec{h}(l(t))$ is time-dependent magnetic field vector in three-dimension with time-dependent control parameter $l(t)$, and $\vec{\mathbf{\sigma}}$ are Pauli matrices. Applying the general formalism discussed above, we obtain the driving Hamiltonian $H ^{\rm{2L}}(t)=H_0 ^{\rm{2L}}(t)+H_1 ^{\rm{2L}}(t)$ \cite{berry2009} with 
\begin{equation}
	H_1 ^{\rm{2L}}(t) = \frac{1}{2|\vec{h}(l)|^2}\left(\vec{h}(l) \times \partial_l {\vec{h}}(l) \right)\cdot \vec{\mathbf{\sigma}}.\label{H1_2L}
\end{equation}

Inspired by the proposals in \cite{del2012assisted} as well as the above simple example of a two-level system, we now turn our attention to the family of $d$-dimensional free-fermion Hamiltonians in the canonical form:
\begin{equation}\label{free_fermion_hamiltonian}
	H_0 ^{\rm{ff}}= \sum_\mathbf{k} \psi_\mathbf{k} ^\dagger \left[\vec{h}_\mathbf{k}(l(t))\cdot \vec{\sigma}_\mathbf{k} \right]\psi_\mathbf{k},
\end{equation}
where $\vec{\sigma}_\mathbf{k}$ denote the Pauli matrices acting on the $\mathbf{k}$-mode and $\psi_\mathbf{k} ^\dagger =\left(c_{\mathbf{k},1}^\dagger,c_{\mathbf{k},2}^\dagger \right)$ are fermionic operators. It was shown in \cite{dziarmaga2005} that the one-dimensional quantum Ising model in a transverse field can be represented by series of independent Landau-Zener transitions, as seen in (\ref{free_fermion_hamiltonian}). Hence, it is apparent that one can generalize the two-level system treatement above to a many-body system, in search of an auxillary Hamiltonian to attain shortcut to adiabaticity. The counterdiabatic term for the quantum Ising model has been found in \cite{del2012assisted,damski2014counterdiabatic}, which can be written in the free-fermion representation (\ref{free_fermion_hamiltonian})-
\begin{equation}
	H_1 ^{\rm{ff}}=l'(t) \sum_{\mathbf{k}} \frac{1}{2\epsilon^2 _\mathbf{k}} \psi_\mathbf{k} ^\dagger \left[\left( \vec{h}_\mathbf{k}(l) \times \partial_l \vec{h}_\mathbf{k}(l) \right) \cdot \vec{\sigma}_\mathbf{k} \right]\psi_\mathbf{k}.
\end{equation}
We will use this formula later to derive the counterdiabatic driving Hamiltonian to achieve cluster states. Here, the instantaneous eigenstates of $H_0 ^{\rm{ff}}$ have associated eigenenergies $\epsilon_{\mathbf{k},\pm}=\pm |\vec{h}_\mathbf{k} (l)|=\pm\sqrt{[h^x _\mathbf{k}(l)]^2 +[h^y _\mathbf{k}(l)]^2 +[h^z _\mathbf{k}(l)]^2}$.

\section{Counterdiabatic driving of 1-D Kitaev honeycomb model}\label{Kitaev_honeycomb}
\subsection{The model}
\begin{figure}[t]
\centering
\includegraphics[scale=0.3]{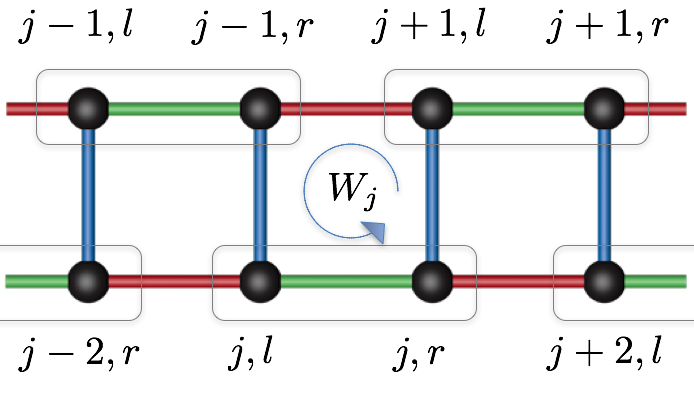}
\caption{Deformed Kitaev honeycomb model in one dimension \cite{kyaw2014measurement}. A grey retangle with two spin-1/2 particles connected by a green bond represents a logical qubit. Black spheres represent spin-1/2 particles, physical qubits, red bonds denote $\sigma^x \sigma^x$, blue ones denote $\sigma^y \sigma^y$, and green ones are $\sigma^z \sigma^z$ interactions, respectively. Here, the subscript $j$ represent locations of the logical qubits within the lattice. $W_j$ is a conserved quantity within each plaquette $j$. Refer to the main text for its expression. We note that there are two notations for each lattice site. One is the position of the physical qubit, which is indicated by blue coloured font within parentheses. For example, {\color{blue}$(i,1)$} indicates a qubit at position $i$ lower row. Another is the location of two ends of the logical qubit, represented by black coloured font. For instance, $j,l$ means left end of the logical qubit $j$. Each bond has distance $a$.}
\label{fig:1Dmodel}
\end{figure}

Let us now put forward everything we have discussed so far within the general framework and apply to a particular model we proposed in \cite{kyaw2014measurement} as shown in figure \ref{fig:1Dmodel}. We emphasize that our proposal to go beyond adiabatic evolution to attain cluster states, is not limited to the example model we present here. It is valid and applicable as long as a system fulfills the criteria listed in section \ref{sec:adiabatic} and it can be represented in the canonical form (\ref{free_fermion_hamiltonian}). 

The 1-D model considered here (see figure \ref{fig:1Dmodel}) arises from an exact analytical analysis made by Kitaev \cite{kitaev2006} in the context of topological quantum computation with anyons. The model Hamiltonian is given by $H_0 ^{\rm{1D}}=H_s ^{\rm{1D}} +\lambda V ^{\rm{1D}}$, with
\begin{eqnarray}\label{1D_hamiltonian}
	H_s ^{\rm{1D}} &=& -J \sum_j \sigma_{j,l} ^z \sigma_{j,r} ^z ,\\
	V ^{\rm{1D}} &=& -\sum_j \left(\sigma_{j,l} ^x \sigma_{j-2,r} ^x + \sigma_{j,l} ^y \sigma_{j-1,r} ^y \right).
\end{eqnarray}
Here, $\sigma^{x,y,z}$ are usual Pauli operators for spin-1/2 particles. The subscript represents location of the physical spin in terms of the logical qubit location. One can also write the Hamiltonian in terms of physical spin coordinates (see figure \ref{fig:1Dmodel}). Each logical qubit, denoted by a grey rectangle, is composed of a pair of spin-1/2 particles. We note that there exists a conserved quantity for each plaquette $j$:
\begin{equation}\label{stabilizer_operators}
	W_j = \sigma_{j,l}^x \sigma_{j,r}^x \sigma_{j-1,r}^z \sigma_{j+1,l}^z , 
\end{equation}
which means $[H_0 ^{\rm{1D}},W_j]=0$, for $\forall j$. Since $W_j$'s commute with each other, they can be diagonalized simultaneously with eigenvalues $w_j = \pm1$, thus allowing us to partition the total Hilbert space into invariant subspaces of $H_0 ^{\rm{1D}}$. A unique ground state with a finite energy gap exists in the subspace with $w_j =+1$, $\forall j$ for $0<\lambda < J/2$, also known as the vortex-free subspace. One can also find similar one-dimensional spins model in \cite{pedrocchi2012,chesi2013vortex}. The Hamiltonian can be diagonalized exactly by first fermionizing the model via a two-dimensional Jordan-Wigner transformation \cite{chen2008exact}:
\begin{eqnarray}
	\sigma^{+}_{\blue{(i, \nu)}}&=& 2\left[\prod_{\nu' <\nu}\prod_{i'}\sigma^z _{\blue{(i' ,\nu')}} \right]\left[\prod_{i' <i}\sigma^z _{\blue{(i' ,\nu)}} \right]c^\dagger _{\blue{(i,\nu)}}, \\
	\sigma^z _{\blue{(i ,\nu)}}&=& 2c^\dagger _{\blue{(i ,\nu)}} c_{\blue{(i ,\nu)}}-1,
\end{eqnarray}
where $\nu=1,2$, since we have only two rows in y direction, and $c_{\blue{(i\nu)}}$ are fermion operators. When $\nu=1$, $\left[\prod_{\nu' <\nu}\prod_{i'}\sigma^z _{\blue{(i' ,\nu')}} \right]= {I}$.  From here onwards, we adopt the notations used in figure \ref{fig:1Dmodel}, where locations with respect to logical qubits are denoted by black coloured font without any parentheses, while locations with respect to physical qubits are represented by blue coloured font inside parentheses. The beginning of the pseudo 1-D chain is at $1,l \equiv \blue{(1,1)}$ and $1,r \equiv \blue{(2,1)}$, for consistency. Subsequently, we apply on-site Majorana fermions transformation
\begin{eqnarray}
	A_l &=& (c+c^\dagger)_l, \hspace{0.5cm} B_l =(c-c^\dagger)_l /i \\
	A_r &=& (c-c^\dagger)_r /i , \hspace{0.5cm} B_r =(c+c^\dagger)_r.
\end{eqnarray}
It is then followed by assigning a fermion in each z-bond as 
\begin{equation}
	d_{j}=(A_{j,r}+i A_{j,l})/2, \hspace{0.5cm} d_{j+1}=(A_{j+1,r}+i A_{j+1,l})/2, 
\end{equation} 
and so on. For sufficiently large systems, the system ground state is at the bulk voltex-free configurations \cite{kitaev2006,chen2008exact,pedrocchi2012}, where $w_j=+1, \forall j$. The exact solution for ground state can be found by a Fourier transformation \cite{chen2008exact,kyaw2015skirting}. The four different transformations, namely the Jordan-Wigner transformation, the on-site Majorana transformation, the z-bond fermion assignment and the Fourier transformation that we have described above, are generically termed as transformation from this point onwards.

As the aim of this section is to obtain the shortcut Hamiltonian for creation of cluster states in the 1-D model (figure \ref{fig:1Dmodel}), we would limit ourselves by considering the periodic boundary condition within the model, rather than considering the infinite chain as in \cite{chen2008exact,kyaw2015skirting}. A few comments are in order. Since we are mainly interested in the vortex-free gapped ground state subspace, we would mainly focus on $w_j=+1$ condition at all times. We will not discuss on any richness of the model. When we apply the periodic boundary condition, there are two ways to do it. One way is to simply connect the two last ends in each row with the two beginning ends as such $\sigma_{\color{blue}{(N_p +1,2)}}=\sigma_{\color{blue}{(1,2)}}$ (the upper row) and $\sigma_{\color{blue}{(N_p +1,1)}}=\sigma_{\color{blue}{(1,1)}}$ (the lower row). However, in this ring-shaped setup, we need that $N_p$, which is the total number of physical spins present in each row, to be an even number. In this way, we obtain one more additional logical qubit from $\sigma^z _{N_p,l}\sigma^z _{Np,r}$. There is another possibility to link up the four extreme ends together by twisting them up like a M\"obius strip. For instance, $\sigma_{\color{blue}{(N_p +1,2)}}=\sigma_{\color{blue}{(1,1)}}$ and $\sigma_{\color{blue}{(N_p +1,1)}}=\sigma_{\color{blue}{(1,2)}}$, but with the requirement that $N_p$ has to be odd. In these manners, we would guarantee to be at the vortex-free subspace as long as $w_j=+1, \forall j$. In this manuscript, we will focus on the former case where $N_p$ is even.

Our model in figure \ref{fig:1Dmodel} is pseudo one-dimensional. Since there are only two sites along $\hat{e}_y$, there are only two allowed momenta $k_y$ in the reciprocal space, which would eventually contribute as a constant during the Fourier transformation. Along the $\hat{e}_x$ direction, we consider the periodic boundary condition with even $N_p$ at each row, which constrains the allowed momenta in the x-direction, with $k_x =q = 0,\pm 2\pi/N_p ,\pm 4\pi/N_p,...,\pm \pi$. After the transformation, the 1-D model Hamiltonian can be written in the free-fermion representation as
\begin{eqnarray}
	H_0 ^{\rm{1D}} (t)= \sum_{q} \psi^\dagger _q &&\left[(J+\lambda(t) (\cos 2{q}a +\cos {q}a))\sigma^z _q \right. \nonumber \\
	&& -\left. i\lambda(t) (\cos 2{q}a +\cos {q}a)\sigma^y _q \right]\psi_q.
\end{eqnarray}
where $\psi^\dagger _q=\left(d^\dagger _q, d_{-q} \right)$ are fermionic operators, and $a$ is the distance between any bond, respectively. Here, we emphasize that $\lambda$ is the only time-dependent parameter that we are tuning during the time evolution, and it comes from the physical Hamiltonian (\ref{1D_hamiltonian}).

By following the previous discussion presented in section \ref{sec:shortcut}, the auxillary Hamiltonian in the reciprocal space is 
\begin{equation}\label{H_1_1d_q}
	H_1 ^{\rm{1D}}(t) = iJ \lambda' (t) \sum_q \mathcal{M}(q) \left[d^\dagger _q d^\dagger _{-q} + d _{-q}d _q \right],
\end{equation}
where
$
	\mathcal{M}(q) = [\cos 2qa +\cos qa]/[J^2 +2 \lambda^2 (\cos^2 2qa +\cos^2 qa +2 (\cos 2qa) (\cos qa))+2J\lambda(\cos 2qa +\cos qa)].
$ 
What we have to do next is to perform the inverse operation of the transformation. As seen from (\ref{H_1_1d_q}), performing the inverse Fourier transformation from the momentum space to the real space spin coordinates is not a straightforward task due to the $q$-dependence in $\mathcal{M}(q)$. However, any periodic function can be expanded in Fourier series of sine and cosine terms. After applying the complete orthogonality relationships of the sine and cosine functions, and the inverse Fourier transformation, we arrive at 
\begin{eqnarray}\label{shortcut_H_logical}
	H_1 ^{\rm{1D}}(t) &=& iJ \lambda' (t) \sum_{j=1}^{N_p} \sum_{n=1}^{N_p} \frac{a_n (\lambda)}{2}\left[ d_{j}^\dagger d^\dagger _{j+n} +d_{j}^\dagger d^\dagger _{j-n} + \textrm{h.c.}  \right]\nonumber\\
	&=&  iJ \lambda' (t) \sum_{j} \sum_{n} a_n (\lambda) \left[d_{j}^\dagger d^\dagger _{j+n} + \textrm{h.c.} \right]. \hspace{0.2cm}\textrm{(periodic boundary)}
\end{eqnarray} 
Here, 
\begin{equation}
	a_n (\lambda) = \frac{1}{N_p} \sum_q \mathcal{M}(q) \cos(n q).
\end{equation}
The result we obtain here in the logical qubit coordinate is the same as \cite{del2012assisted}. However, since our model is pseudo one-dimensional, we arrive at more complicated shortcut Hamiltonian, in the physical spin coordinate. Its expression is dependent on two factors: 1) the location of $j$th logical qubit, whether it is at the lower or upper row, and 2) whether $n$ is odd or even. For the case when $n$ is even, we have a reflection symmetry between the lower and upper rows. That is
\begin{eqnarray}
\fl
	\left[d_{j}^\dagger d^\dagger _{j+n} + \textrm{h.c.} \right]=-\frac{1}{4}&& \left[ \sigma^x _{\blue{(j+1,R)}}\left(\prod_{j+1 < i'<j+n} \sigma^z _{\blue{(i',R)}} \right) \sigma^x _{\blue{(j+n,R)}} \right.\nonumber \\
	  && \left. -\sigma^y _{\blue{(j,R)}} \left(\prod_{j < i' <j+n+1} \sigma^z _{\blue{(i',R)}} \right) \sigma^y _{\blue{(j+n+1,R)}} \right], \textrm{with }R=1,2.
\end{eqnarray}
For the case when $n$ is odd, due to a translational invariance and the commutation relations of the Pauli matrices, we have 
\begin{eqnarray}
\fl
	\left[d_{j}^\dagger d^\dagger _{j+n} + \textrm{h.c.} \right]_{\vec{x}}=-\frac{1}{4}&& \left[ \sigma^x _{\blue{(j+1,1)}}\left(\prod_{j+1 < i' \leq N_p} \sigma^z _{\blue{(i',1)}} \right) \left(\prod_{1\leq i' <j+n} \sigma^z _{\blue{(i',2)}} \right)\sigma^x _{\blue{(j+n,2)}} \right.\nonumber \\
	  && \left. -\sigma^y _{\blue{(j,1)}} \left(\prod_{j < i' \leq N_p} \sigma^z _{\blue{(i',1)}} \right) \left(\prod_{1\leq i' <j+n+1} \sigma^z _{\blue{(i',2)}} \right)\sigma^y _{\blue{(j+n+1,2)}} \right]\nonumber\\
	  &&= e^{i\pi} \left[d_{j}^\dagger d^\dagger _{j+n} + \textrm{h.c.} \right]_{\vec{x}\pm\vec{\tau}},
\end{eqnarray}
where the subscript vector represents the location of the $j$th logical qubit along the chain where it can be located either at the lower or upper row. Upon translation in $\pm \vec{\tau}$ along $\hat{e}_x$, the $j$th qubit is now at the different row as compared to before the translation. Due to this result, in the summation of (\ref{shortcut_H_logical}), when $j$th and $j+n$th logical qubits are located in different rows, the sum of $(d^\dagger _j d^\dagger _{j+n}+d^\dagger _{j+n} d^\dagger _j + \textrm{h.c.})$ would equal to $0$. Therefore, we only need to consider the terms with any two logical ones locating at the same row. In the following subsection, we provide numerical evidence of the benefit of involving shortcut Hamiltonian (\ref{shortcut_H_logical}) to obtain cluster states in our proposed 1-D model.

\subsection{Numerical results}
First, we like to emphasize that there is a distinct difference between the standard adiabatic quantum computing (AQC) scheme, which has been extensively studied in the context of adiabatic shortcut \cite{torrontegui2013} and the one we are implementing here. In AQC, there is a unique ground state at the start of evolution, where a physical system is prepared. This ground state is then adiabatically moved to another unique ground state, by slowly varying some system parameters. To be more elaborate and appropriate to the present discussion, let us take an example of the quantum Ising spin chain in transverse magnetic field. There are two distinct phases in this model namely ferromagnetic and paramagnetic ones. The system is initially prepared in one of the phases and swapped through a quantum critical point to reach to the another phase. Since the system energy gap closes at the quantum critical point, one needs extensive $M$-body counterdiabatic driving terms in order to follow the adiabatic trajectory \cite{del2012assisted,damski2014counterdiabatic}. 

\begin{figure}[t]
\centering
\includegraphics[scale=0.4]{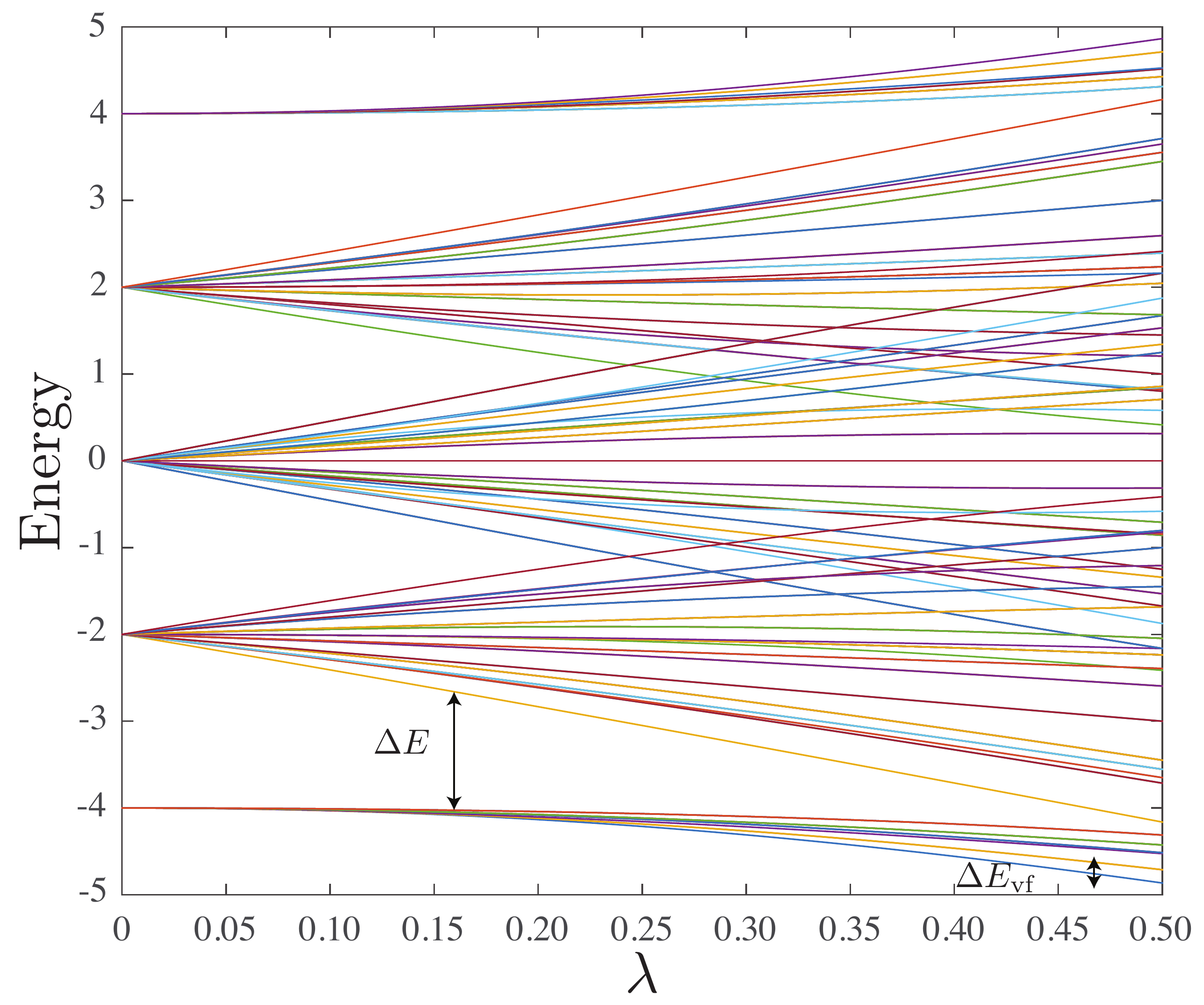}
\caption{Energy spectrum of the deformed Kitaev honeycomb model in 1-D is plotted against the coupling $\lambda$. We consider 8 physical spins with the ring-shaped periodic boundary condition. Please refer to the main text for more detailed information on this particular boundary condition we consider. There are total of $2^8=256$ eigenstates of which 16 of them are belonged to the vortex-free subspace. Within that subspace, the target cluster state, which is also one of the degenerate ground states, is located near $\lambda\approx0.$}
\label{fig:Kitaev_ring_spectrum}
\end{figure}    
Here, in our study, the 1-D model always remains in the same vortex-free subspace $(w_j =+1), \forall j$ throughout the evolution. At the start $\lambda=0.5 [J]$, the system has a unique ground state with energy gap $\Delta E_{\rm{vf}}$ (see figure \ref{fig:Kitaev_ring_spectrum}). As mentioned earlier, that ground state is not a cluster state. In order to get cluster states, we are proposing to cool the system down to its unique ground state and switch off $\lambda$ adiabatically which is the couplings between any $\sigma^x \sigma^x $ and $\sigma^y \sigma^y$ bond. As seen in the figure \ref{fig:Kitaev_ring_spectrum}, the system energy gap $\Delta E_{\rm{vf}}\rightarrow 0$ when $\lambda$ is being turned off. Therefore, there is no unique ground state at the end. In addition, there is also no level crossing present throughout, due to the presence of large $\Delta E$, and we intend to stop the evolution at near the critical point $\lambda\approx 0$.

\begin{figure}[t]
\centering
\includegraphics[scale=0.6]{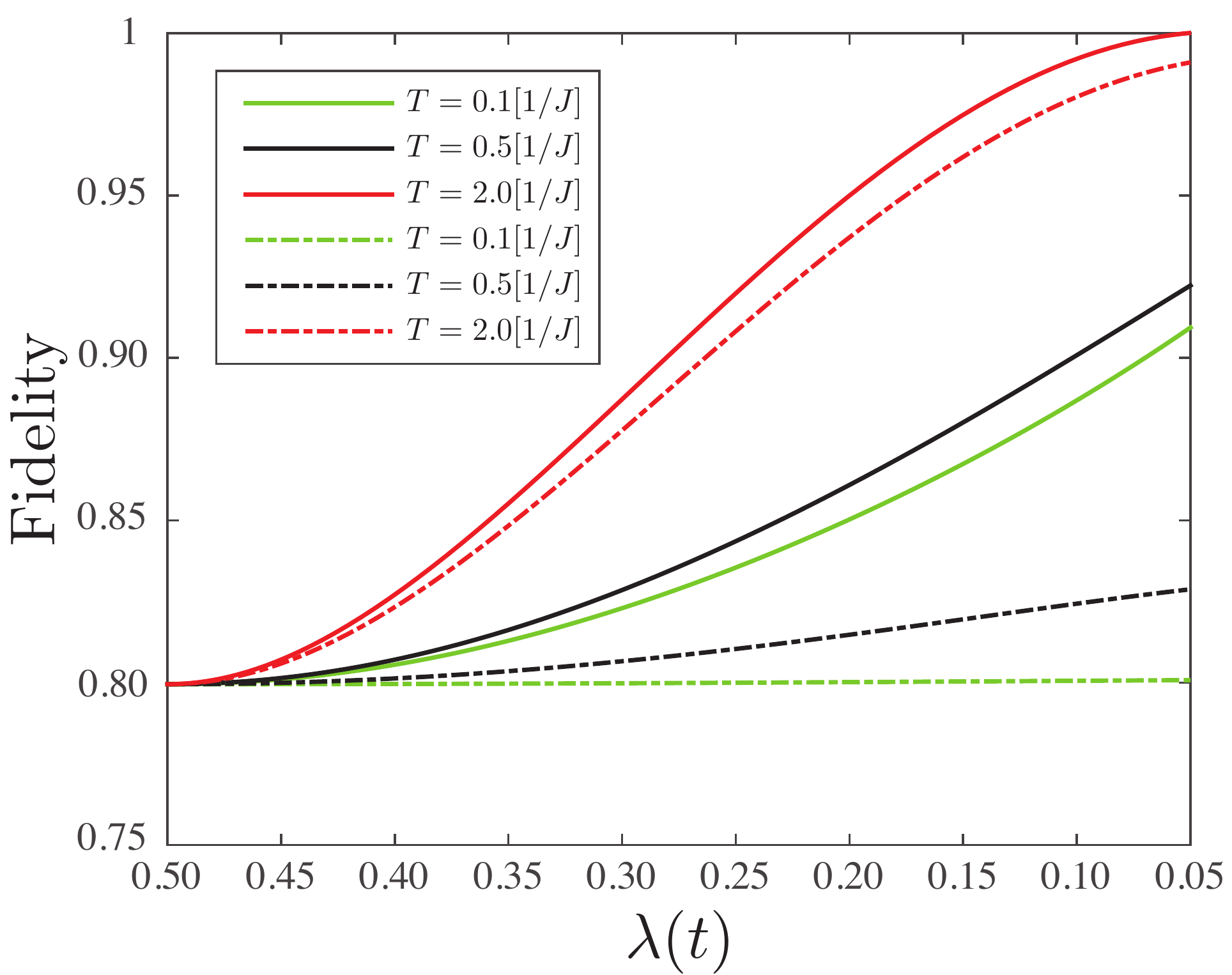}
\caption{Fidelity of cluster state generation for various total evolution time, with or without the counterdiabatic driving Hamiltonian. All the solid lines are the time evolution under the counterdiabatic driving Hamiltonian while the dash-dotted lines correspond to the time evolution under the original one. Fidelity enhancement can be seen with the involvement of the shortcut Hamiltonian. In-depth explanation can be found in the main text.}
\label{fig:cluster_state_fidelity}
\end{figure}

Secondly, the stabilizers $W_j$'s (\ref{stabilizer_operators}), which commute with the system original Hamiltonian, are available throughout the evolution. More importantly, $[V^{\rm{1D}},W_j]=0, \forall j$. That means the instantaneous ground state is protected by the set of stabilizers. Because of this, one does not need to slow down the evolution even though $\Delta E_{\rm{vf}}\rightarrow 0.$ Alternatively, we note that the initially prepared ground state is an eigenstate of the stabilizers. Since $[V^{\rm{1D}},W_j]=0, \forall j$, any change in $V^{\rm{1D}}$ has no effect in the instantaneous ground state. The state will remain in the same one as Bythe beginning. Thus, the slow-down is not necessary. In the numerical simulation shown in figure \ref{fig:cluster_state_fidelity}, we consider total of 8 physical spins in the ring-shaped periodic boundary condition. The quench dynamics is governed by $\lambda(t)=\lambda_0 + 3(\lambda_f -\lambda_0)(t/T)^2 -2(\lambda_f -\lambda_0)(t/T)^3$, with the boundary conditions: $\lambda(0)=\lambda_0$, $\lambda(T)=\lambda_f$ and $\dot{\lambda}(0)=\dot{\lambda}(T)=0$, which is inspired by \cite{saberi2014adiabatic} and references therein.

In the figure \ref{fig:cluster_state_fidelity}, we plot the fidelity between the desired cluster state and the instantaneous eigenstate from the time evolution $F=|\braket{\psi_{\rm gf}}{\psi(t)}|^2$, given that the initial state is the unique ground state. In order to obtain the cluster state $\ket{\psi_{\rm gf}}$ numerically, we diagonalize sum of all the four stabilizer operators and take the 16 eigenstates with eigenvalues $+4$ (vortex-free) among 256 eigenstates. By calculating the fidelity with all 16 states for a very long evolution time $T$, we find the exact numerical expression of the desired cluster state, since it is the only state that gives fidelity $1$. This expression of the eigenstate is then used for all the plots seen in the figure. 

As discussed before, any variation in $\lambda$ does not excite the initial ground state to any other subspaces. Does it imply one can arbitrarily quench the original Hamiltonian within a short time $T$ to obtain the cluster state, without invoking the shortcut Hamiltonian? As it turns out, if the quench rate is too fast, the initial state remains almost stationary even when $\lambda\rightarrow 0$. And, the state does not evolve out of the initially prepared subspace that is the lowest energy band as seen in figure \ref{fig:Kitaev_ring_spectrum}. This resultant effect can be clearly seen from the green dash-dotted line in the figure \ref{fig:cluster_state_fidelity} when we consider the total evolution time of $0.1 [1/J]$. With relatively longer time evolution, one approaches the desired cluster state as the fidelity value tends to $1$, which can be seen from black and red dash-dotted lines in the figure. With the inclusion of the shortcut Hamiltonian we found in (\ref{shortcut_H_logical}), the fidelity is enhanced for all the three previous evolutions as seen from the solid lines. For the red coloured solid line, we surprisingly observe that we can already achieve the target state even at $\lambda=0.05 [J]$.

\section{Conclusion}\label{conclusion}
In this manuscript, we show that one can obtain cluster states which are the resources for the measurement-based quantum computing via shortcut to adiabaticity. Our proposal is inspired by the recent development in adiabatic tracking of quantum many-body dynamics \cite{del2012assisted,damski2014counterdiabatic,saberi2014adiabatic,mukherjee2016local} and it is then combined with the technique \cite{kyaw2014measurement,kyaw2015skirting} we developed to skirt around the no-go theorem \cite{nielsen2006} in the universal MBQC. With the inclusion of the shortcut Hamiltonian, the fidelity to obtain cluster states is enhanced. However, the resultant shortcut Hamiltonian is of the form obtained in \cite{del2012assisted}, with the requirement of $M$-body interacting Hamiltonians. Hence, it is still hard to realize experimentally. It will be nice to develop a mean to reduce the complexity of the auxillary Hamiltonian by optimising control parameters as similar to \cite{saberi2014adiabatic}, which we will leave for the future work. 

We also remark that one-dimensional cluster state is not a resource state for the universal MBQC. However, it can be used to achieve an arbitrary single-qubit gate \cite{raussendorf2001} or a quantum wire \cite{gross2007}. Eventually, we hope to get resource states beyond this one-dimensional model via transitionless quantum driving algorithm. However,  the auxillary Hamiltonians for the 2-D and 3-D Hamiltonians presented in \cite{kyaw2014measurement} appears vastly complicating with our current technique. In light of recent advancement in quantum computing experiments, we believe that our proposed model, the original Hamiltonian, can easily be realized with state-of-the-art superconducting circuit architecture since what we need is nearest-neighbour two-body Ising interaction \cite{bernien2017,roushan2017}. The auxillary Hamiltonian with long-range interactions required for the adaibatic shortcut could also be achieved with cavity mediated qubit-qubit interaction \cite{kyaw2015creation}.

\section*{Acknowledgements}
We are grateful to Adolfo del Campo for his very useful comments and feedback, which improves the quality of the manuscript. T.H.K also acknowledges helpful discussions and feedback from Victor M. Bastidas, Guillermo Romero, Beno\^it Gr\'emaud, and Shabnam Safaei. The funding support from the National Research Foundation \& Ministry of Education, Singapore, is acknowledged.


\end{document}